\documentclass[9pt,twocolumn,twoside]{opticajnl}
\journal{opticajournal} 
\usepackage{siunitx}
\setboolean{shortarticle}{true}
\newcommand{\dif}{\mathrm{d}}
\newcommand{\myrev}[1]{#1}
\newcommand{\myrevv}[1]{#1}
\newcommand{\myscale}{0.88}

\usepackage{lineno}

\title{Statistical mechanics and pressure of composite multimoded weakly nonlinear optical systems}

\author[1,2,*]{Nikolaos K. Efremidis}
\author[3]{Demetrios N. Christodoulides}

\affil[1]{Department of Applied Mathematics, University of Crete, Heraklion 71409, Greece}
\affil[2]{Institute of Applied and Computational Mathematics, FORTH, 70013 Heraklion, Crete, Greece}
\affil[3]{Ming Hsieh Department of Electrical and Computer Engineering, University of Southern California, Los Angeles, California 90089, USA}
\affil[*]{nefrem@uoc.gr}

\begin{abstract}

  Statistical mechanics can provide a versatile theoretical framework for investigating the collective dynamics of weakly nonlinear waves-settings that can be utterly complex to describe otherwise. In optics, composite systems arise due to interactions between different frequencies and/or polarizations. The purpose of this work is to develop a thermodynamic theory that takes into account the synergistic action of multiple components.
  We find that the type of the nonlinearity involved can have important implications in the thermalization process and, hence, can lead to different thermal equilibrium conditions. 
  Importantly, we derive closed-form expressions for the actual optomechanical pressure that is exerted on the system. In particular, the total optomechanical pressure is the sum of the partial pressures due to each component. Our results can be applied to a variety of weakly nonlinear optical settings such as multimode fibers, bulk waveguides, photonic lattices, and coupled microresonators. We present two specific examples, where two colors interact in a waveguide array with either a cubic or quadratic nonlinearity.
  
\end{abstract}

\setboolean{displaycopyright}{false} 

\begin{document}

\maketitle


There is a variety of multimoded optical settings such as multimode fibers~\cite{richa-np2013}, multicore fibers, waveguide arrays~\cite{chris-nature2003}, as well as cavity and microcavity arrays in the time domain~\cite{vahal-nature2003}.
The complexity associated with spatial or temporal multimoded nonlinear arrangements is such that on most occasions it makes it virtually impossible to either understand or predict their behavior. Hence, developing a comprehensive theory would be of paramount importance in investigating their collective behavior. Recently a thermodynamic formalism based on supermode dynamics was put forward~\cite{wu-np2019}.

From the perspective of applications, high-power delivery using large area multimode fibers is particularly appealing for mode-locking~\cite{wrigh-science2017} and supercontinuum generation~\cite{wrigh-np2015,lopez-ol2016}.
In this respect, the unexpected phenomenon of ``beam self-cleaning'' was observed in graded index fibers~\cite{lopez-ol2016,liu-ol2016,krupa-np2017}: as the power of the laser beam increases the beam profile at the output turns from a highly speckled pattern to a bell shaped beam. This effect has a thermodynamic origin, and does not depend on the kind of the nonlinearity involved involved, as long as four-wave mixing processes ensue~\cite{pourb-np2022,mangi-oe2022}.

Furthermore, attractive or repulsive optomechanical forces between suspended coupled waveguides lead to measurable displacement of their separation~\cite{povin-ol2005,povin-oe2005}, resulting to possible applications ranging from optical routing~\cite{rosen-np2009} and optical information storage~\cite{huang-acs2019}, to precision measurements~\cite{anets-np2009}, photothermal sensing~\cite{prues-acs2018}, and actuators~\cite{li-np2009,roels-nn2009,ren-ACSNano2013}. Importantly, for more complicated arrangements involving a large number of waveguides, resonators, or more generally supermodes, utilization of the optical thermodynamic theory leads to
the calculation of the applied pressure and, thus, to the computation of the exerted optomechanical forces~\cite{efrem-cp2022}.

Recently, thermal equilibrium in inverted distributions associated with negative temperatures have been observed~\cite{muniz-science2023,baudin-prl2023}.
\myrevv{The computation of the temperature and the chemical potential as a function of the extensive system parameters was analyzed in}~\cite{parto-ol2019}, and in~\cite{makri-ol2020,ramos-prx2020} a grand canonical approach was utilized. Extensivity, differential relations, and the effect of additional system parameters were examined in~\cite{efrem-pra2021}. This theory was utilized for the computation of the optomechanical pressure~\cite{efrem-cp2022}. Kinetic nonequilibrium theories can also be used to analyze the process of thermalization~\cite{pitois-prl2006,lagra-el2007,suret-prl2010}. 

Nonlinear processes involving different colors and/or polarizations give rise to a number of interesting and important phenomena with applications in harmonic generation, parametric amplification, and optical communications~\cite{agrawal-nfo}.
The understanding of the statistical properties and the applied optomechanical pressure of such systems is the main focus in this work. We find that the type of nonlinearity involved can affect the resulting thermal equilibrium conditions. Specifically, the fundamental parameters are the power conservation laws of the system, each one being associated with a different chemical potential. We derive expressions for the resulting Rayleigh-Jeans distributions, the internal energy and its differential, and a Gibbs-Duhem equation. In addition, we calculate the optomechanical pressure exerted by each component. We find that the total pressure is the direct sum of these partial pressures. We present two particular examples of composite systems in a discrete one-dimensional (1D) lattice. Specifically, we examine two color interactions under (a) self- and cross-phase modulation, and (b) quadratic nonlinearity between the fundamental and the second harmonic.

\begin{figure}[t!]
  \centering
  \includegraphics[width=\linewidth]{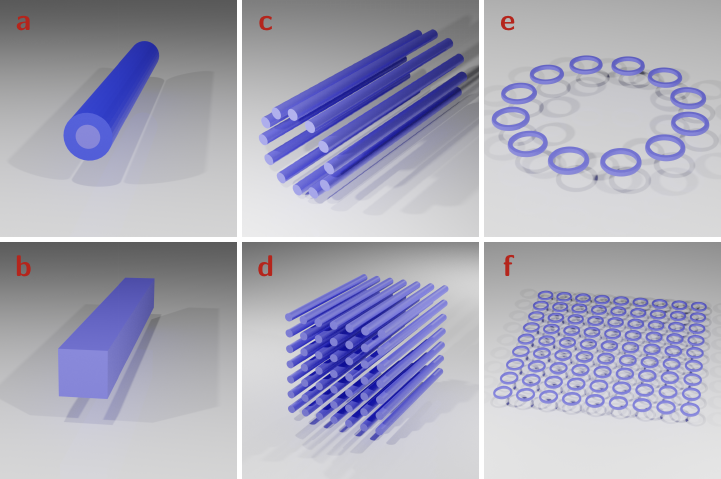}  
  \caption{Multimoded optical settings where the optical thermodynamic theory of composite systems can be applied, include bulk and discrete systems in space or time domain. Typical examples are (a) multimode fibers (b) multimode waveguides, (c) 1D and (d) two-dimensional (2D) waveguide arrays, and (e) 1D and (f) 2D coupled ring resonators.}
  \label{fig:01}
\end{figure}

Let us consider an optical system, \myrev{typical examples are depicted in Fig.~\ref{fig:01}}, with $\gamma=1,\ldots,\Gamma$ components that may represent different frequencies and/or polarizations.
Without any loss of generality, and for the sake of simplicity in our notation, we make two simplifications: First, although the number of modes $M_\gamma$ might be different for each component, we assume that $M_\gamma=M$. Second, we restrict ourselves to 1D arrangements. 
The $\gamma$ component of the optical wave is expanded in modal space as
 $ |\psi_\gamma\rangle=\sum_{l=1}^MC_\gamma^{(l)}(z)|\psi_\gamma^{(l)}\rangle, $
where the supermodes \myrev{(the modes of the total system)} of the same component are considered to be orthonormal $\langle\psi_\gamma^{(l)}|\psi_\gamma^{(l')}\rangle=\delta_{l,l'}$.
The wave propagates along the $z$-direction, however, similar calculation can be carried out in the time domain.
The power and the internal energy (which physically represents the energy per unit of propagation length) of the $\gamma$ component are given by
\begin{equation}
  N_\gamma= \sum\nolimits_{l=1}^{M}n_\gamma^{(l)}\myrev{(z)},\quad
  \label{eq:Ngamma}
\end{equation}
\begin{equation}
  U_\gamma=
  \sum\nolimits_{l=1}^M\varepsilon_\gamma^{(l)}n_\gamma^{(l)}\myrev{(z)}/\omega_\gamma,
  \label{eq:Ugamma}
\end{equation}
where $n_\gamma^{(l)}=|C_\gamma^{(l)}|^2$ \myrev{and $\varepsilon_\gamma^{(l)}$} is the power \myrev{and the propagation constant, respectively,} of the $l$th mode of the $\gamma$ component and $\omega_\gamma$ is the frequency of the $\gamma$ component. 

The first conservation law is the internal energy of the system, which is the sum of the linear energies of each component along with an additional nonlinear term that accounts for interactions
\begin{equation}
  U_\mathrm{tot} =
  U+U_\mathrm{int}=
  \sum\nolimits_{\gamma=1}^\Gamma U_\gamma+U_\mathrm{int}. 
  \label{eq:U}
\end{equation}
The type of the nonlinearity has qualitative implications in the thermalization process:
Depending on $U_\mathrm{int}$, the system can possess different power conservation laws which, in turn, affect the resulting thermal equilibrium. 
To make our notation simpler, and without loss of generality, we consider two particular cases which we will separately study:
(i) Each component $\gamma$ maintains its power $N_\gamma$ resulting to $\Gamma$ power conservation laws and (ii) only the total power is conserved
\begin{equation}
  N = \sum\nolimits_{\gamma=1}^\Gamma N_\gamma.
  \label{eq:N}
\end{equation}

We start from the first case where the power conservation laws are given by $N_\gamma$, $\gamma=1,\ldots,\Gamma$ [Eq.~(\ref{eq:Ngamma})]. We follow an approach similar to Ref.~\cite{efrem-pra2021} that takes into account the additional system parameters. We \myrev{maximize} the Gibbs entropy in phase space
under the constraints given by the $M+1$ conservation laws \myrev{of Eqs.~(\ref{eq:Ngamma}), (\ref{eq:U})}. The resulting probability is
$\rho(\{n_\gamma^{(l)}\}) = \exp\left[
  -q-\sum\nolimits_{\gamma}\alpha_\gamma N_\gamma-\beta U
\right]$
where $\alpha_\gamma$, $\beta$ are the Lagrange multipliers, and the $q$-potential
\begin{equation}
  q =
  \sum\nolimits_\gamma q_\gamma =
  -\sum\nolimits_{l,\gamma}
  \log[\alpha_\gamma+\beta\varepsilon_\gamma^{(l)}/\omega_\gamma],
  \label{eq:q}
\end{equation}
\myrev{which is related to the grand canonical partition function $Q$ through $q=\log Q$,} is the normalization constant so that $\int\rho\prod_{\gamma,l}\dif n_\gamma^{(l)}=1$. 
The average value of a quantity $r$ \myrev{in thermal equilibrium} is given by
$\overline r = \int r\rho\prod_{\gamma,l}\dif n_\gamma^{(l)}.$
Thus, we find 
\begin{equation}
  \overline n_\gamma^{(l)} = -\frac{\omega_\gamma}\beta
  \left(
    \frac{\partial q}{\partial\varepsilon_\gamma^{(l)}}
  \right)_{\alpha,\beta,M,\overline\varepsilon_\gamma^{(l)}}
  =\frac1{\alpha_\gamma+\beta\varepsilon_\gamma^{(l)}/\omega_\gamma},
  \label{eq:overnl}
\end{equation}
\begin{equation}
  \overline N_\gamma = - 
  \left(
    \partial q/\partial\alpha_\gamma
  \right)_{\overline\alpha_\gamma,\beta,M,\xi}
  =\sum\nolimits_{l=1}^{M}\overline n_\gamma^{(l)},
  \label{eq:overN}
\end{equation}
and
\begin{equation}
  \overline U = - 
  (
    \partial q/\partial\beta
  )_{\alpha,\beta,M,\xi}
  =\sum\nolimits_{\gamma,l}\overline n_\gamma^{(l)}\varepsilon_\gamma^{(l)}/\omega_\gamma,
  \label{eq:overU}
\end{equation}
where $\overline\varepsilon_\gamma^{(l)}=\{\varepsilon_1^{(1)},\ldots,\varepsilon_\Gamma^{(M)}\}\backslash\varepsilon_\gamma^{(l)}$ and $\overline\alpha_\gamma=\{\alpha_1,\ldots,\alpha_\Gamma\}\backslash\alpha_\gamma$. From this point on, since we are mainly interested in ensemble average values, for simplicity we omit the overline notation. We take the differential of $q(\alpha_\gamma,\beta,s,M)$, where in the case of one transverse direction $L$ is the transverse size (length) of the system,
and define the intensive parameter $s=L/M$. Note that for two transverse directions we need to define two such  variables, one for each transverse direction. Following calculations similar to those of Refs.~\cite{efrem-pra2021,efrem-cp2022} we obtain the following equation for 
\begin{equation}
  \dif U = T\dif S -pM\dif s -R\dif M +\sum\nolimits_\gamma\mu_\gamma\dif N_\gamma.
  \label{eq:difU}
\end{equation}
In Eq.~(\ref{eq:difU}), $T$ is the \myrev{ optical temperature (not to be confused with the environmental temperature)} and $\mu_\gamma$ is the chemical potential of the $\gamma$ component. These parameters are related to the Lagrange multipliers via $\alpha_\gamma=-\mu_\gamma/T$ and $\beta=1/T$.
\myrevv{The modal occupancies can be expressed in terms of $T$ and $\mu_\gamma$ as} 
\begin{equation}
  n_\gamma^{(l)} =
  T/(\varepsilon_\gamma^{(l)}/\omega_\gamma-\mu_\gamma).
  \label{eq:nl}
\end{equation}
In Eq.~(\ref{eq:difU})
\begin{equation}
  R =
   T\left(
     \partial q/\partial M
  \right)_{T,\{\mu_\gamma\},s}
  \label{eq:RR}
\end{equation}
is the internal pressure that describes how the internal energy is modified when an additional mode is added. Importantly,
\begin{equation}
  p_\gamma = 
  -\sum_{l=1}^{M_\gamma}
  \frac{n_\gamma^{(l)}}{\omega_\gamma}
  \left(
    \frac{\partial\varepsilon_\gamma^{(l)}}{\partial L}
  \right)_M
  \label{eq:pgamma}
\end{equation}
is the actual optomechanical pressure that is exerted on the system due to the $\gamma$-component of the optical wave. The total pressure is the sum of the partial pressures of the components
\begin{equation}
  p =\sum\nolimits_{\gamma=1}^\Gamma p_\gamma.
\end{equation}
From Eqs.~(\ref{eq:overnl})-(\ref{eq:overU}) we derive the following equation of state that relates the energy and the power conservation laws with the optical temperature and the chemical potentials
\begin{equation}
  U - \sum\nolimits_{\gamma=1}^\Gamma\mu_\gamma N_\gamma = \Gamma M T. 
  \label{eq:state}
\end{equation}
Note that the entropy is directly related to the $q$-potential via
\begin{equation}
  S = q + \Gamma M.
  \label{eq:S}
\end{equation}

If we assume that the entropy $S$ is extensive with respect to $(U,\{N_\gamma\},M)$, meaning that $S (\lambda U,\{\lambda N_\gamma\},\lambda M)=\lambda S(U,\{N_\gamma\},M)$, then we can directly integrate Eq.~(\ref{eq:difU}) leading to
\begin{equation}
  U = TS -RM+\sum\nolimits_\gamma\mu_\gamma N_\gamma.
  \label{eq:Uext}
\end{equation}
In addition, from Eqs.~(\ref{eq:state}), (\ref{eq:S}), (\ref{eq:Uext}), we derive
\begin{equation}
  q = R M/T.
  \label{eq:qqq}
\end{equation}
From Eqs.~(\ref{eq:RR}), (\ref{eq:qqq}), we see that for extensive systems
\begin{equation}
  \left(
    \frac{\partial q_\gamma}{\partial M}
  \right)_{T,\{\mu_\gamma\},s}
  =
  \frac{q_\gamma}{M}.
  \label{eq:qoverM}
\end{equation}
By combining Eq.~(\ref{eq:difU}) with Eq.~(\ref{eq:Uext}), we obtain a Gibbs-Duhem equation that relates variations of the intensive parameters
\begin{equation}
  S\dif T + \sum\nolimits_\gamma N_\gamma\dif\mu_\gamma+pM\dif s -M\dif R = 0. 
  \label{eq:GD}
\end{equation}

In case (ii) only the total power is conserved. We define $\alpha$ to be the Lagrange multiplier of $N$. The calculations are similar to case (i), and thus, we focus on explaining how the previous results are modified. In particular, in Eqs.~(\ref{eq:q}), (\ref{eq:overnl}), (\ref{eq:overN}), (\ref{eq:overU}), we have to replace $\alpha_\gamma\rightarrow\alpha$. Since $\alpha=-\mu/T$, by substituting $\mu_\gamma\rightarrow\mu$ to Eq.~(\ref{eq:nl}) we find
\begin{equation}
  n_\gamma^{(l)} = \frac{T}{\varepsilon_\gamma^{(l)}/\omega_\gamma-\mu}.
  \label{eq:nl2}
\end{equation}
The differential of the internal energy [Eq.~(\ref{eq:difU})], and the equation of state [Eq.~(\ref{eq:state})] take the form
\begin{equation}
  \dif U = T\dif S -pM\dif s -R\dif M +\mu\dif N,
  \label{eq:difU2}
\end{equation}
and
\begin{equation}
  U - \mu N = \Gamma M T. 
  \label{eq:state2}
\end{equation}

If, in addition, the system is extensive, then the internal energy can be expressed as
\begin{equation}
  U = TS -RM+\mu N.
  \label{eq:Uext2}
\end{equation}
As a result we find the following Gibbs-Duhem equation
\begin{equation}
  S\dif T + N\dif\mu+pM\dif s -M\dif R = 0. 
  \label{eq:GD2}
\end{equation}
Finally Eq.~(\ref{eq:qoverM}) holds as is. 

Note that our calculations are system-independent and are expected to apply to any type of composite multimoded optical systems, such as those depicted in Fig.~\ref{fig:01}. However, if we assume that the propagation constants have the functional form
\begin{equation}
  \varepsilon_\gamma^{(l)}=\varepsilon_{\gamma,0}+
  \kappa(s)G(\gamma,l)
  =
  \varepsilon_{\gamma,0}+
  \tilde\varepsilon_\gamma^{(l)}(\kappa)
  \label{eq:varepsilon_discrete}
\end{equation}
where $\varepsilon_{\gamma,0}$ are constants, then
\begin{equation}
  \left(
    \frac{\partial\varepsilon_\gamma^{(l)}}{\partial L}
  \right)_M
  =
  \frac1M\left(
    \frac{\partial\varepsilon_\gamma^{(l)}}{\partial s}
  \right)_M
  =
  \frac1M
  \frac{\tilde\varepsilon_\gamma^{(l)}}{\kappa}
  \frac{\dif\kappa}{\dif s}.
\end{equation}
As a result, the pressure due to the $\gamma$ component in Eq.~(\ref{eq:pgamma}) takes the simpler form
\begin{equation}
  p_\gamma =
  -\frac{\tilde U_\gamma}{M}
  \frac{\dif\log\kappa_\gamma}{\dif s},
  \label{eq:pgamma2}
\end{equation}
where $\tilde U_\gamma$ represents a shifted value of the internal energy
\begin{equation}
  \tilde U_\gamma =
  \sum_{l=1}^M
  \tilde\varepsilon_\gamma^{(l)}n_\gamma^{(l)}/\omega_\gamma, 
  \label{eq:tildeU}
\end{equation}
such that $U_\gamma=\tilde U_\gamma+N_\gamma\varepsilon_{\gamma,0}/\omega_\gamma$.
Note that Eq.~(\ref{eq:varepsilon_discrete}) can describe for example monoatomic discrete systems, such as waveguide arrays and coupled resonators. 
In our previous work, we have considered the pressure in single component systems~\cite{efrem-cp2022}, and we have set the energy offset $\varepsilon_\gamma$ to zero leading to $\tilde U=U$. Here, for generality, we opt to use the unshifted values of the propagation constants. As expected, the energy shifts $\varepsilon_{\gamma,0}$, do not modify the thermodynamic results (such as the optical temperature and the Rayleigh-Jeans distribution) but only affect the resulting values of the chemical potentials, which are also shifted accordingly.

\begin{figure}[t!]
  \centering
  \includegraphics[width=\myscale\linewidth]{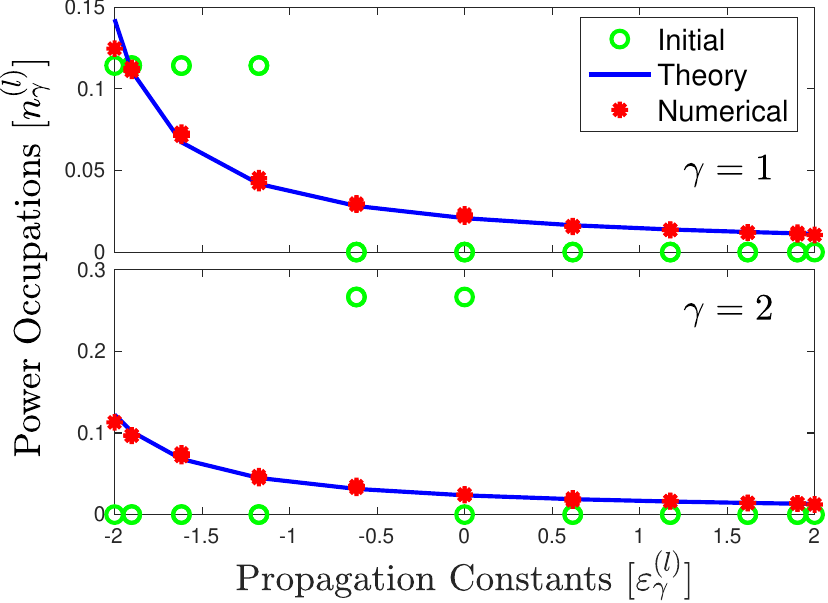}  
  \caption{Thermalization in a two-color discrete optical system with a cubic nonlinearity described by Eqs~(\ref{eq:cubic1})-(\ref{eq:cubic2}).
    The initial power occupation numbers are shown with  green circles. The theoretical prediction [based on Eq.~(\ref{eq:nl})] is depicted with the blue line and the numerical averages are shown with  red stars. 
  }
  \label{fig:cubic}
\end{figure}
We next examine particular settings where our theory can be applied. For the purposes of this study, we limit ourselves to the case of photonic lattices. As a first example, we consider a system with a cubic nonlinearity and two different colors. The discrete coupled-mode theory equations for a 1D waveguide lattice are given by
\begin{align}
  i\dot\phi_m
  +\kappa_1(\phi_{m+1}+\phi_{m-1})
  + 
  \omega_1\sigma
  (|\phi_m|^2+
  2|\psi_m|^2)\phi_m
  & =0, \label{eq:cubic1} \\
  i\dot\psi_m
  +\kappa_2(\psi_{m+1}+\psi_{m-1})
  + 
  \omega_2\sigma
  (|\psi_m|^2+
  2|\phi_m|^2)\psi_m
  & =0, \label{eq:cubic2}
\end{align}
where $\phi_m$, $\psi_m$ are the field components, $m=1,\ldots,M$ is the waveguide (lattice) site, $\kappa_\gamma$ are the coupling coefficients, $\sigma$ is related to the strength of the nonlinearity, and $\omega_\gamma$ are the optical frequencies. In addition to the Hamiltonian, Eqs.~(\ref{eq:cubic1})-(\ref{eq:cubic2}) have two power conservation laws, one for each color $N_1=\sum_m|\phi_m|^2$, $N_2=\sum_m|\psi_m|^2$. The supermode propagation constants in the case of periodic boundary conditions are given by
\begin{equation}
  \varepsilon_\gamma^{(l)}=-2
  \myrevv{\kappa_\gamma(s)}\cos(2\pi l/M),\quad l=1,\ldots,M
  \label{eq:propconst}
\end{equation}
and, thus, follow the functional form of Eq.~(\ref{eq:varepsilon_discrete}) with $\varepsilon_{\gamma,0}=0$ meaning that $\tilde U_\gamma=U_\gamma$.

We select an array with $M=20$ waveguides and use dimensionless normalized parameters $\omega_1=1$, $\omega_2=1.2$, $\kappa_1=\kappa_2=1$, and $\sigma=1$. \myrev{We numerically solve Eqs.~(\ref{eq:cubic1})-(\ref{eq:cubic2}) using Runge-Kutta methods.} As shown in Fig.~\ref{fig:cubic}, we initially equally excite all the modes with energies between $-2\le\varepsilon_1^{(l)}\le-1$ and \myrevv{$-1\le\varepsilon_2^{(l)}\le0$}, for the first and second component, respectively, with the same amount of power. The modal phases are randomly selected using a uniform distribution. Since $N_1=N_2=0.8$, the initial energies are then given by $U_1(z=0)=-1.3$, $U_2(z=0)=-0.27$. The theoretical prediction for the ensemble average values in thermal equilibrium are $T=0.0487$, $\mu_1=-2.3413$, $\mu_2=-2.0638$, $U_1=-0.9$, and $U_2=-0.68$.
In Fig.~\ref{fig:cubic}, we see that we have excellent agreement between the theoretically predicted Rayleigh-Jeans distribution and the simulated average values of the power occupation numbers. 
Using Eq.~(\ref{eq:tildeU}) we can compute the optomechanical pressure exerted on the system due each color $p_1=0.045\dif\kappa_1/\dif s$, $p_2=0.034\dif\kappa_1/\dif s$. The total pressure is the sum of these two partial pressures.

\begin{figure}[t!]
  \centering
  \includegraphics[width=\myscale\linewidth]{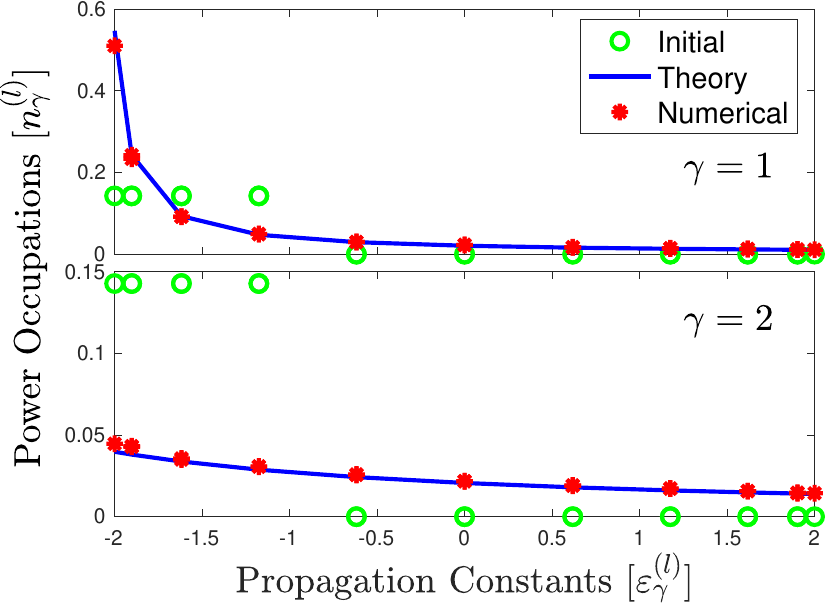}
    \caption{Thermalization in a two-color discrete optical system with a quadratic nonlinearity as described by Eqs~(\ref{eq:SHG1})-(\ref{eq:SHG2}).
    The initial power occupation numbers are shown with green circles. The theoretical prediction, based on Eq.~(\ref{eq:nl2}), is depicted with the blue line and the numerical averages are show with red stars.}
  \label{fig:quadratic}
\end{figure}
As a second example, we consider a 1D discrete lattice with a quadratic nonlinearity (second harmonic generation) and periodic boundary conditions. The coupled-mode theory equations that describe this system are given by
\begin{align}
  i\dot\phi_m
  +\kappa_1(\phi_{m+1}+\phi_{m-1})
  + 
  \omega_1\sigma\psi_m\phi_m^*
  & =0, \label{eq:SHG1} \\
  i\dot\psi_m
  -\Delta\beta\psi_m
  +\kappa_2(\psi_{m+1}+\psi_{m-1})
  + 
  (\omega_2/2)\sigma\phi_m^2
  & =0. \label{eq:SHG2}
\end{align}
Here, restrict our attention to phase-matched conditions $\Delta\beta=0$, in which case the propagation constants reduce to those of Eq.~(\ref{eq:propconst}). The main difference from the first example, is that power can flow between the two colors $\omega_1$ and $\omega_2=2\omega_1$. Thus, only the total power $N=\sum_m(|\phi_m|^2+|\psi_m|^2)$ is conserved.

We consider a lattice with $M=20$ sites, and for both colors, we excite all the modes that satisfy $-2\le\varepsilon_\gamma^{(l)}\le-1$, $\gamma=1,2$ with the same amount of power and a uniformly random phase [see Fig.~\ref{fig:quadratic}]. In particular, we have $N_1=N_2=1$ that leads to an initial energy distribution $U_1(z=0)=-1.63$ and $U_2=-0.81$. Our theoretical calculations show that for these values of energy and power, at thermal equilibrium the system should have an optical temperature $T=0.043$ and a common chemical potential $\mu=-2.1$. Furthermore, the average thermal equilibrium values of the power and energy are $N_1=1.53$, $N_2=0.47$, $U_1=-2.32$, $U_2=-0.12$. We see that most of the power is directed towards the fundamental harmonic. The optomechanical pressure exerted by each component, as computed by Eq.~(\ref{eq:pgamma2}), is $p_1=0.116\dif\kappa_1/\dif s$, $p_2=0.006\dif\kappa_1/\dif s$. In Fig.~\ref{fig:quadratic}, we depict the theoretically predicted averages of the power occupation numbers $n_\gamma^{(l)}$, which are in excellent agreement with the numerically calculated average values after the system has reached thermal equilibrium. 

In conclusion, we have theoretically examined the statistical mechanics of weakly nonlinear \myrev{(in the sense that nonlinearity is mainly used to achieve wave mixing)} composite multimoded optical systems. Our calculations, lead to expressions for the main thermodynamic variables including the exerted optomechanical pressure. These results can be applied to a variety of multimoded optical settings including bulk and discrete arrangements in space and time-domain.

\begin{backmatter}
  
  \bmsection{Funding}
  The work of DC was partially supported by: Israel Ministry of Defense (4441279927); MPS Simons Collaboration (733682); Department of Energy (DE-SC002228); Army Research Office (W911NF-23-1-0312); Office of Naval Research (N00014- 20-1-2789); Air Force Office of Scientific Research (FA9550-20-1-0322).

  \bmsection{Disclosures} The authors declare no conflicts of interest.

  \bmsection{Data Availability Statement} Data underlying the results presented in this paper are not publicly available at this time but may be obtained from the authors upon reasonable request.

\end{backmatter}

\newcommand{\noopsort[1]}{} \newcommand{\singleletter}[1]{#1}

\cleardoublepage

\renewcommand\refname{FULL REFERENCES}

\end{document}